\documentclass{PoS}
\pdfoutput=1
\usepackage{rotating,booktabs,amsmath,amssymb, multicol,url}
\usepackage[utf8]{inputenc}
\usepackage[numbers,sort&compress]{natbib}
\usepackage{natbibspacing}
\usepackage{ulem}

\title{Semi-leptonic form factors for $B_s \to K \ell \nu$ and $B_s \to D_s \ell \nu$ }

\ShortTitle{Semi-leptonic form factors for $B_s \to K \ell \nu$ and $B_s \to D_s \ell \nu$ }

\author{Jonathan M.~Flynn\\
  Physics and Astronomy, University of Southampton, Southampton SO17 1BJ, UK}

\author{Ryan C.~Hill\\
  Physics and Astronomy, University of Southampton, Southampton SO17 1BJ, UK\newline
  DISCnet Centre for Doctoral Training, University of Southampton, Southampton SO17 1BJ, UK}

\author{Andreas J\"uttner\\
Physics and Astronomy, University of Southampton, Southampton SO17 1BJ, UK}

\author{Amarjit Soni\\
  Physics Department, Brookhaven National Laboratory, Upton, NY 11973, USA}

\author{Justus Tobias Tsang\\
Higgs Centre for Theoretical Physics, The University of Edinburgh, EH9 3FD, UK}

\author{\speaker{Oliver Witzel}\\
        Department of Physics, University of Colorado Boulder, Boulder, CO, USA\\
        E-mail: \email{Oliver.Witzel@colorado.edu}
}
\abstract{Semi-leptonic $B_s \to K \ell \nu$ and $B_s \to D_s \ell \nu$ decays provide an alternative $b$-decay channel to determine the CKM matrix elements $|V_{ub}|$ and $|V_{cb}|$ or to obtain $R$-ratios to investigate lepton flavor universality violations. In addition, these decays may shed further light on the discrepancies seen in the analysis of inclusive vs. exclusive decays.  Using the nonperturbative methods of lattice QCD, theoretical results are obtained with good precision and full control over systematic uncertainties. This talk will highlight ongoing efforts of the $B$-physics program by the RBC-UKQCD collaboration.}

\FullConference{The 36th Annual International Symposium on Lattice Field Theory - LATTICE2018\\
                22-28 July, 2018\\
                Michigan State University, East Lansing, Michigan, USA.}

\begin{document}
\section{Introduction}

Since the discovery of the Higgs boson in 2012, no other new elementary particles have been discovered nor have direct signs of new physics been detected at the Large Hadron Collider (LHC).  Hence testing the Standard Model (SM) at high precision is increasingly important, with flavor physics (electro-weak interactions changing quark flavor) of special relevance. At tree-level, the SM allows up-type quarks to decay to down-type quarks (and vice versa) by emitting a charged $W^\pm$ boson, but flavor changing neutral currents (FCNC) are suppressed and occur only at loop level. High precision calculations of processes allowed in the SM are needed to test for differences to experimental observations. Deviations could signal new physics, arising for example from virtual particles in loops.

New physics is expected to occur at higher energy scales and observing its effects is more likely if the decaying particle can release large amounts of energy. Decays of mesons containing a heavy $b$-quark provide many opportunities because the b-quark lives long enough for experimental investigation but also delivers more than 4 GeV energy. The large $b$-mass also allows a plethora of decay channels and correspondingly many tests of the SM. $B_{(s)}$-meson decays could allow indirect observation of new physics from virtual particles which would otherwise show up at currently inaccessible energy scales. It is important to perform tests for both tree-level as well as loop-level processes. For both cases tantalizing deviations between SM predictions and experimental measurements have been reported see e.g.~\cite{HFLAV:RdRds-2018,Gershon:2017jlb} and references within. Particularly striking are ratios investigating the universality of lepton flavors in semi-leptonic $B$ decays, e.g., for $B$ mesons decaying to a $D^{(*)}$ meson with either $\tau \nu_\tau$ or $\mu \nu_\mu$ leptons in the final state,
\begin{align}
  R_{D^{(*)}}^{\tau/\mu}\equiv \frac{BF(B\to D^{(*)}\tau\nu_\tau)}{BF(B\to D^{(*)} \mu \nu_\mu)}.
\end{align}
Currently the combined analysis for pseudoscalar and vector hadronic final states yields a tension of more than $3\sigma$ between SM prediction and the experimental values obtained from measurements by BaBar, Belle, and LHCb \cite{Lees:2012xj,Huschle:2015rga,Aaij:2015yra,Sato:2016svk,Hirose:2016wfn,Aaij:2017uff}.

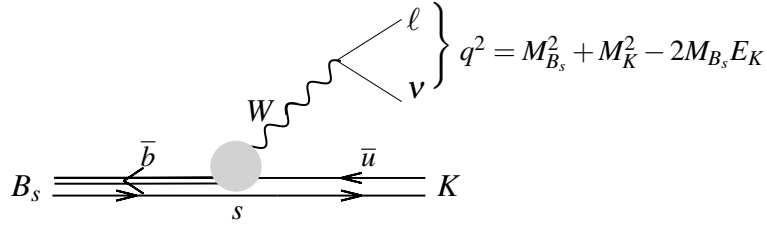
\begin{figure}[tb]
  \centering
  \begin{picture}(100,30)
    \put(6,2){\includegraphics[width=50mm]{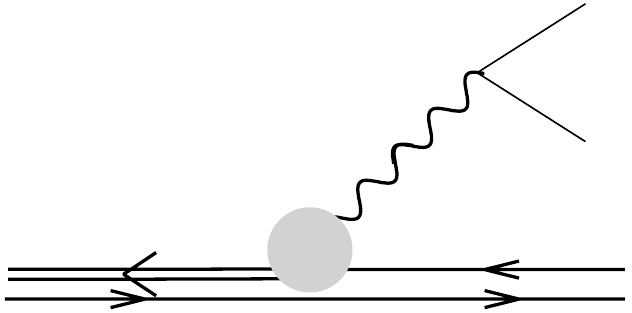}}
    \put(1,3){\large{$B_s$}} \put(57,3){\large{$K$}}
    \put(32,13){\large{$W$}} 
    \put(53,25){\large{$\ell$}} \put(53,16){\large{$\nu$}}
    \put(56,22){$\Bigg\}$}\put(60,21){$q^2= M_{B_s}^2 + M_K^2 - 2 M_{B_s} E_K$}
    \put(30,0){$s$}
    \put(47,7){$\overline{u}$}
    \put(18,7){$\overline{b}$}
  \end{picture}
  \caption{Sketch of tree-level weak semi-leptonic $B_s$ decays mediated by a charged $W^\pm$ boson and a set-up with the $B_s$ meson at rest.}
  \label{fig.treelevel}
\end{figure}

Our project for nonperturbative calculations of semileptonic $B_{(s)}$ decays includes operators for tree- and loop-level processes with one pseudoscalar or one vector hadronic final state \cite{Flynn:2016vej}, but here we focus on tree-level $B_s$-meson decays with a kaon or $D_s$ meson in the final state.  Experimentally many $B_s$ decays are observed by LHCb and we intend to support their program by pursuing these calculations. The diagram corresponding to $B_s\to K \ell \nu$  decays is sketched in Fig.~\ref{fig.treelevel}; for $B_s\to D_s \ell\nu$ decays kaons and $D_s$-mesons are simply exchanged and the $\bar u$ daughter quark becomes $\bar c$. 
Conventionally, the branching fraction measured experimentally is parameterized by
\begin{align}
       \frac{d\Gamma(B_s\to K\ell\nu)}{dq^2} & = \frac{G_F^2 |V_{ub}|^2}{24 \pi^3} \,\frac{(q^2-m_\ell^2)^2\sqrt{E_K^2-M_K^2}}{q^4M_{B_s}^2}
        \bigg[ \left(1+\frac{m_\ell^2}{2q^2}\right)M_{B_s}^2(E_K^2-M_K^2)|f_+(q^2)|^2 \nonumber\\
&+\,\frac{3m_\ell^2}{8q^2}(M_{B_s}^2-M_K^2)^2|f_0(q^2)|^2
          \bigg]\,. \label{eq:B_semileptonic_rate}
%
%
\end{align}
The nonperturbative contributions are given by the form factors $f_+$ and $f_0$ which are related to the matrix element
\begin{align}
  \langle K |V^\mu | B_s\rangle = f_+(q^2) \left( p^\mu_{B_s} + p^\mu_K - \frac{M^2_{B_s} - M^2_K}{q^2}q^\mu\right) + f_0(q^2)  \frac{M^2_{B_s} - M^2_K}{q^2}q^\mu.
\end{align}
The weak decay is dominated by short distance contributions and hence we can consider the weak operator as a point-like object and implement the calculation using conventional lattice QCD techniques. In the following we report updates on our efforts to determine the form factors for $B_s\to K\ell \nu$ and $B_s\to D_s\ell\nu$ decays. Our calculations are based on a subset of RBC-UKQCD's 2+1 flavor domain wall fermion and Iwasaki gauge field ensembles \cite{Allton:2008pn,Aoki:2010dy,Blum:2014tka,Boyle:2017jwu} which we summarize in Tab.~\ref{tab.ensembles}. Light and  strange quarks are simulated using domain wall fermions \cite{Kaplan:1992bt,Shamir:1993zy,Furman:1994ky,Brower:2012vk}, charm quarks are simulated by applying the M\"obius domain wall action to heavy quarks \cite{Boyle:2016imm}, and bottom quarks are simulated using the relativistic heavy quark (RHQ) action \cite{Christ:2006us,Lin:2006ur}, a variant of the Fermilab action \cite{ElKhadra:1996mp} with nonperturbatively tuned parameters \cite{Aoki:2012xaa}. Further details of the set-up and our project to compute bottom and charm physics can be found in Refs.~\cite{Christ:2014uea,Flynn:2015mha,Flynn:2015xna,Boyle:2017jwu,Boyle:2018knm}. Here we focus on updates of our form factor calculations reporting in Section \ref{Sec.BsK} on $B_s\to K \ell\nu$ decays and in Section \ref{Sec.BsDs} on $B_s\to D_s\ell\nu$, before summarizing in Section \ref{Sec.summary}.

\begin{table}[tb]
  \centering
  \begin{tabular}{c@{~~~}c@{~}c@{~}l@{~}l@{~}c@{~}rc}\toprule
&L & $a^{-1}$(GeV) & ~$am_l$ & ~$am_s$ & $M_\pi$(MeV)& \# configs.&\#sources\\ \midrule
C1&24 & 1.784 &  0.005 & 0.040 & 338 & 1636~~~~&1\\
C2&24 & 1.784 &  0.010 & 0.040 & 434 & 1419~~~~&1\\
\midrule
M1&32 & 2.383 &  0.004 & 0.030 & 301 & 628~~~~&2\\
M2&32 & 2.383 &  0.006 & 0.030 & 362 & 889~~~~&2\\
M3&32 & 2.383 &  0.008 & 0.030 & 411 & 544~~~~&2\\
\midrule
F1&48 & 2.774 &0.002144  & 0.02144 & 234 &98~~~~& 24\\
\bottomrule
  \end{tabular}
  \caption{Dynamical 2+1 flavor domain-wall fermion ensembles \cite{Allton:2008pn,Aoki:2010dy,Blum:2014tka,Boyle:2017jwu} used in this calculation. The lattice spacing is determined in combined analysis \cite{Blum:2014tka,Boyle:2017jwu} and the quoted values correspond to $a$ $\sim 0.11$ fm, $\sim 0.08$ fm, $\sim 0.07$ fm.}
  \label{tab.ensembles}
\end{table}

\section{Form factors for semi-leptonic $B_s \to K\ell \nu$ decays}
\label{Sec.BsK}
\begin{figure}[tb]
  \centering
  \includegraphics[height=0.28\textheight]{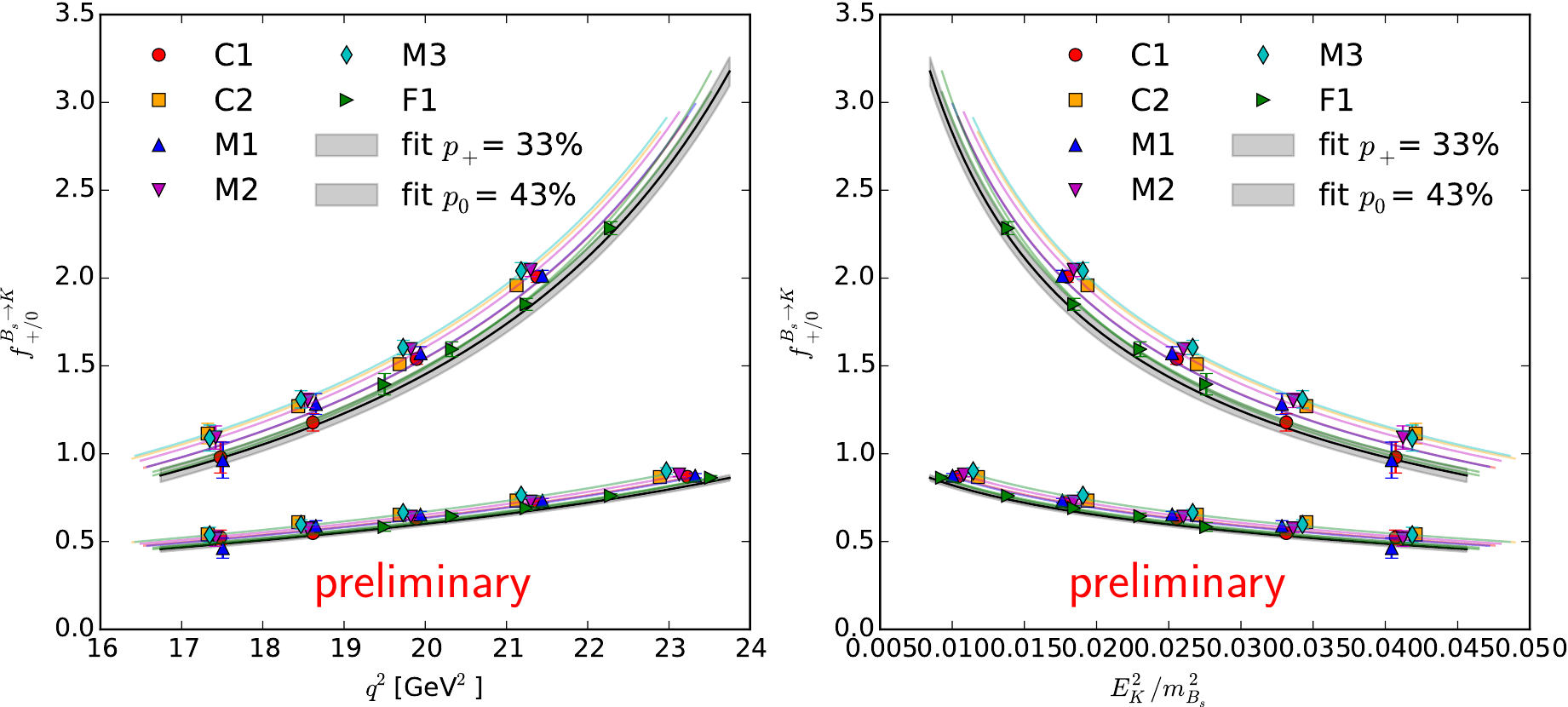}
  \caption{Chiral-continuum extrapolation of semi-leptonic form factors for $B_s\to K \ell\nu$ decays using HM$\chi$PT. Only statistical errors are shown. The plot on the left presents the form factors in units of $q^2$, while the plot on the right uses the kaon energy squared normalized by the $B_s$-meson mass.}
  \label{fig.BsK_formfactors}
\end{figure}

In order to extend our original work \cite{Flynn:2015mha} and include for example the new ensemble $F_1$ at a third, finer lattice spacing, we first repeated our non-perturbative tuning of the RHQ parameters \cite{Aoki:2012xaa} to reflect updated values of the lattice spacing and the physical mass of the strange quark \cite{Blum:2014tka}. Using the newly tuned RHQ parameters, we simulate physical $b$-quarks and use close-to physical values for the strange quark, whereas the mass of light quark is set to the unitary light quark mass on each ensemble. We calculate 3-point and 2-point functions to extract the form factors on each ensemble using discrete spatial lattice momenta up to $p^2= 4 (2\pi/L)^2$. This results in the set of colored data points for $f_+$ and $f_0$ shown in Fig.~\ref{fig.BsK_formfactors}. Using an ansatz based on heavy meson chiral perturbation theory (HM$\chi$PT) \cite{Becirevic:2002sc,Bijnens:2010ws}, we obtain a functional form to describe our data 
\begin{align}
f_\text{pole} (M_K, E_K,a^2)&= \frac{1}{E_K +\Delta} c^{(1)} \cdot\Big[ 1+\frac{\delta f}{(4\pi f)^2}+ c^{(2)}\frac{M_K^2}{\Lambda^2} +  c^{(3)} \frac{E_K}{\Lambda} +  c^{(4)} \frac{E_K^2}{\Lambda^2}+ c^{(5)} \frac{a^2}{\Lambda^2 a^4_{32}} \Big],   
\end{align}
where $\delta f$ are non-analytic logarithms of the kaon mass and the hard-kaon limit is taken by $M_K/E_K\to 0$. Next we perform a global fit to all data points for $f_+$ ($f_0$) to obtain form factors in the chiral-continuum limit. These fits have excellent $p$-values of 33\% for $f_+$ (43\% for $f_0$) and the outcome is shown by the black central line with gray error band in Fig.~\ref{fig.BsK_formfactors}. 

\begin{figure}[tb]
   \includegraphics[height=0.24\textheight]{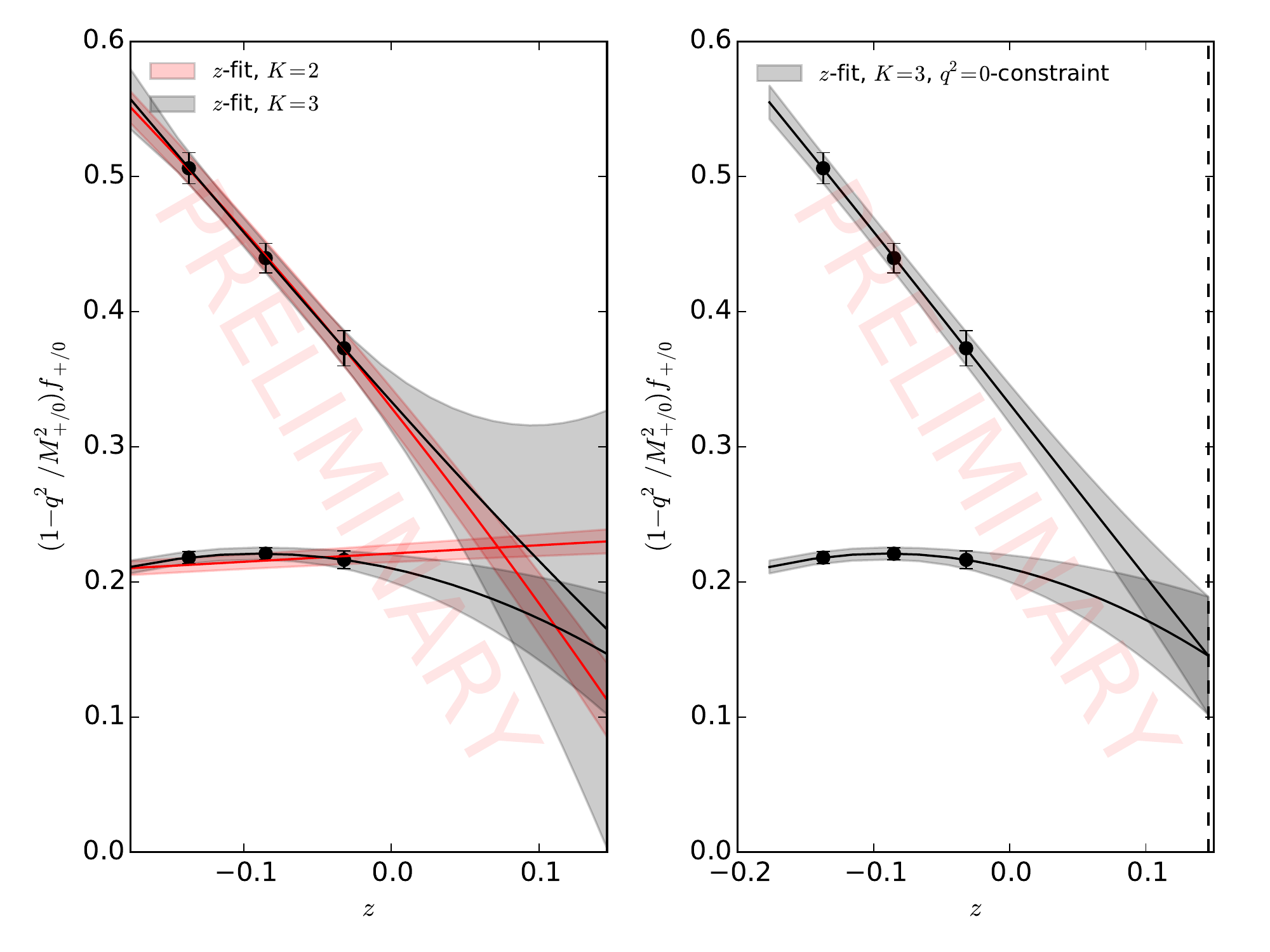} \hfill
   \includegraphics[height=0.24\textheight]{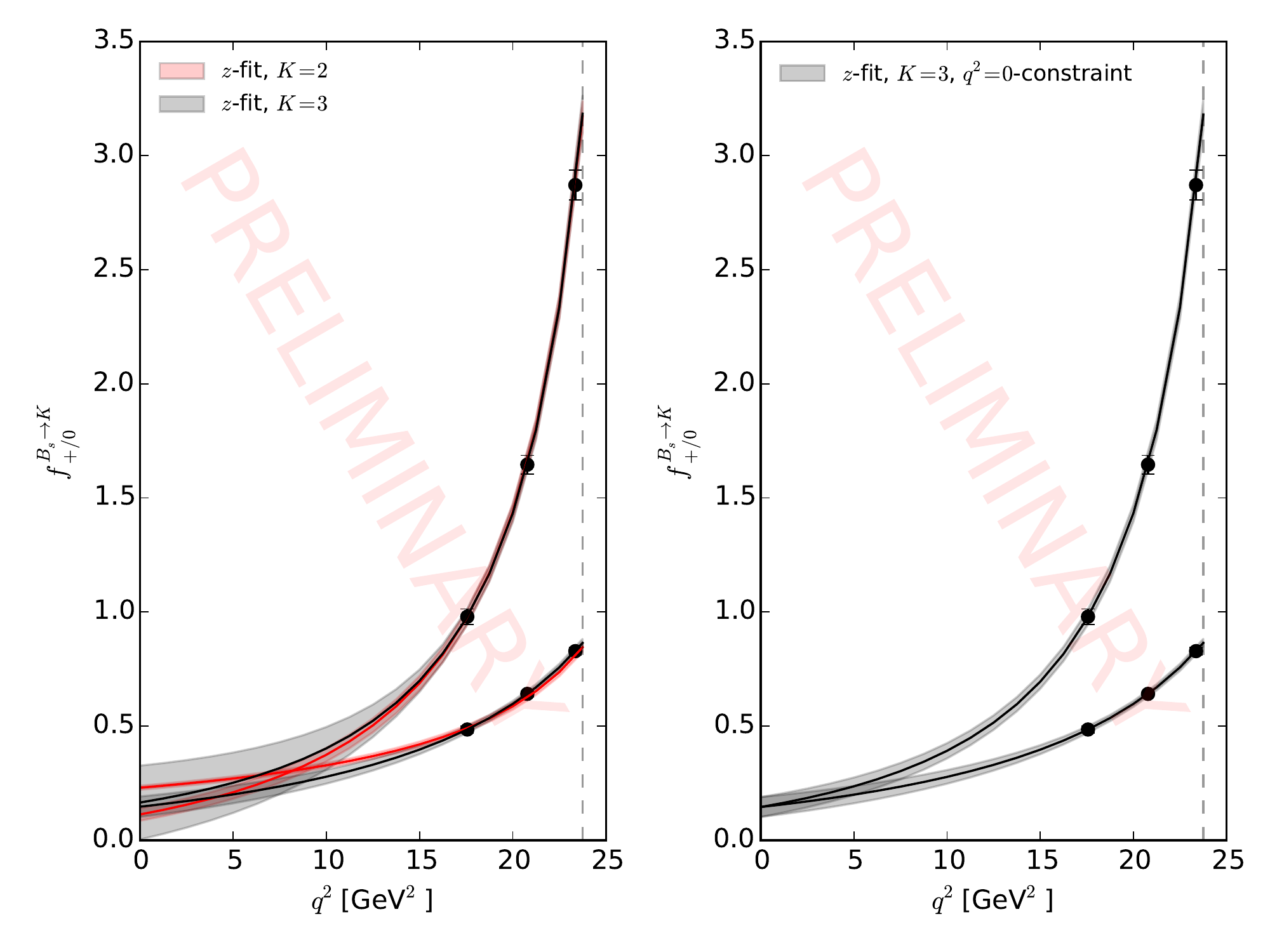}
   \caption{Kinematical extrapolation of the $B_s\to K\ell\nu$ form factors using the BCL $z$ expansion over the full $q^2$ range for $K=2$ or 3. On the left, the result are shown in units of the $z$ extrapolation, on the right in physical GeV\textsuperscript{2}. }
   \label{fig.BsK_zfit}
\end{figure}

We aim for a continuum description of the form factors with a full statistical and systematic error budget. The latter is still work in progress. Thus uncertainties presented in the following are neither final nor complete. Using however our continuum limit result with a preliminary accounting of systematic effects, we can carry out an extrapolation over the entire range of allowed $q^2$ values. We do so by implementing a so called $z$ expansion i.e.~we map $q^2$ to the variable $z$ using 
\begin{align}
  z(q^2,t_0) = \frac{\sqrt{1 -q^2/t_+} - \sqrt{1-t_0/t_+}}{\sqrt{1 -q^2/t_+}+ \sqrt{1-t_0/t_+}}
\end{align}
with $t_\pm = \left(M_{B_s}\pm M_K\right)^2$ and $t_0\equiv t_\text{opt} = (M_{B_s}+M_K)(\sqrt{M_{B_s}}-\sqrt{M_K})^2$.
The outcome of this kinematical extrapolation is presented in Fig.~\ref{fig.BsK_zfit} where we use the parametrization by Bourrely, Caprini, and Lellouch (BCL) \cite{Bourrely:2008za}
\begin{align}
  f_+(q^2) &= \frac{1}{1-q^2/M_{B^*}^2} \sum_{k=0}^{K-1} b_+^{(k)}\left[ z^k -\frac{k}{K}(-1)^{k-K}z^K\right];\qquad
 f_0(q^2) &=  \frac{1}{1-q^2/M_{B^*(0^+)}^2} \sum_{k=0}^{K-1} b_0^{(k)} z^k.
\end{align}  
We show extrapolations of our results using $K=2$ and 3 and in addition also implement the constraint $f_0(0)=f_+(0)$.

\section{Form factors for $B_s \to D_s\ell \nu$ decays}
\label{Sec.BsDs}

\begin{figure}[tb]
    \includegraphics[height=0.25\textheight]{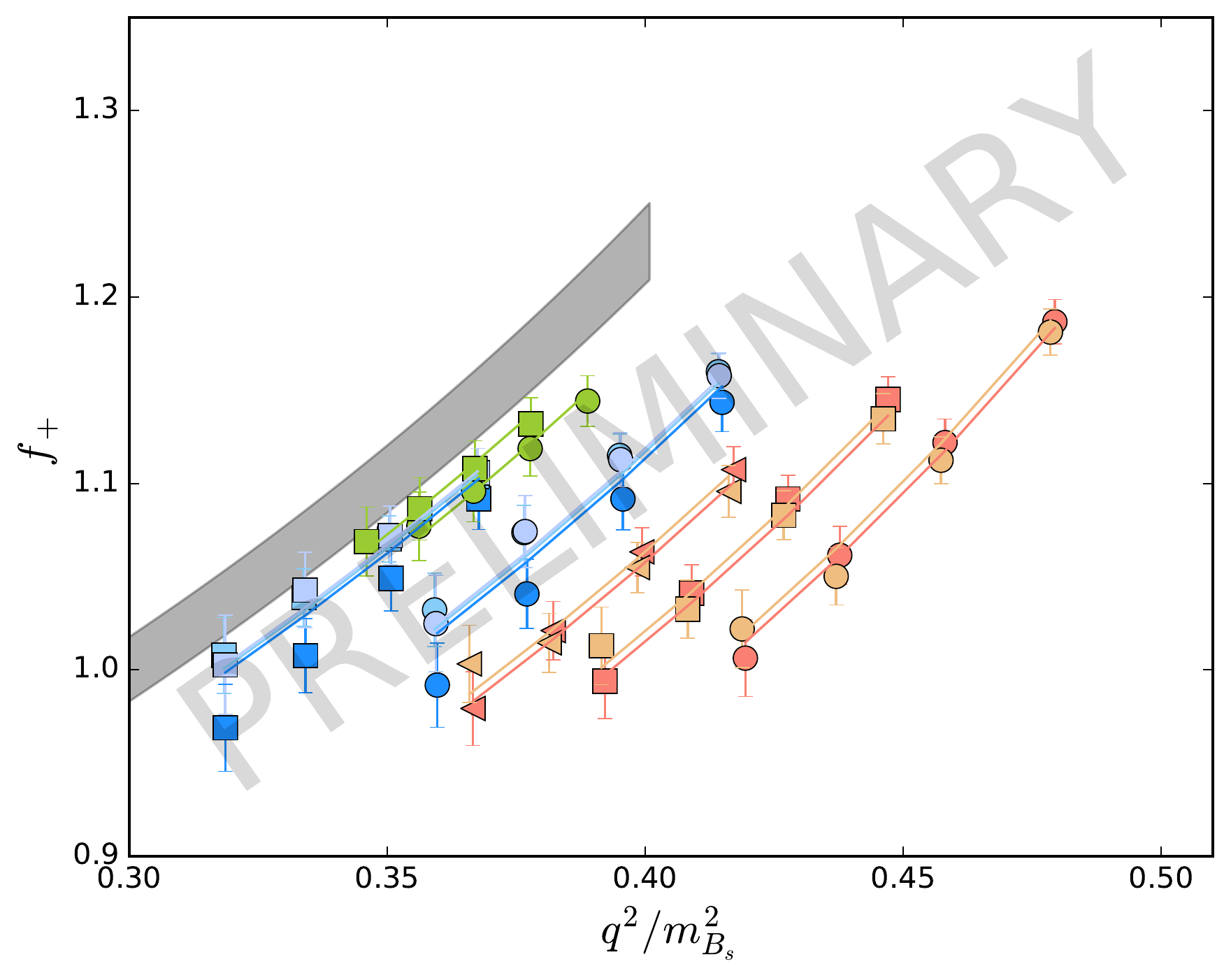}  \hfill
    \includegraphics[height=0.25\textheight]{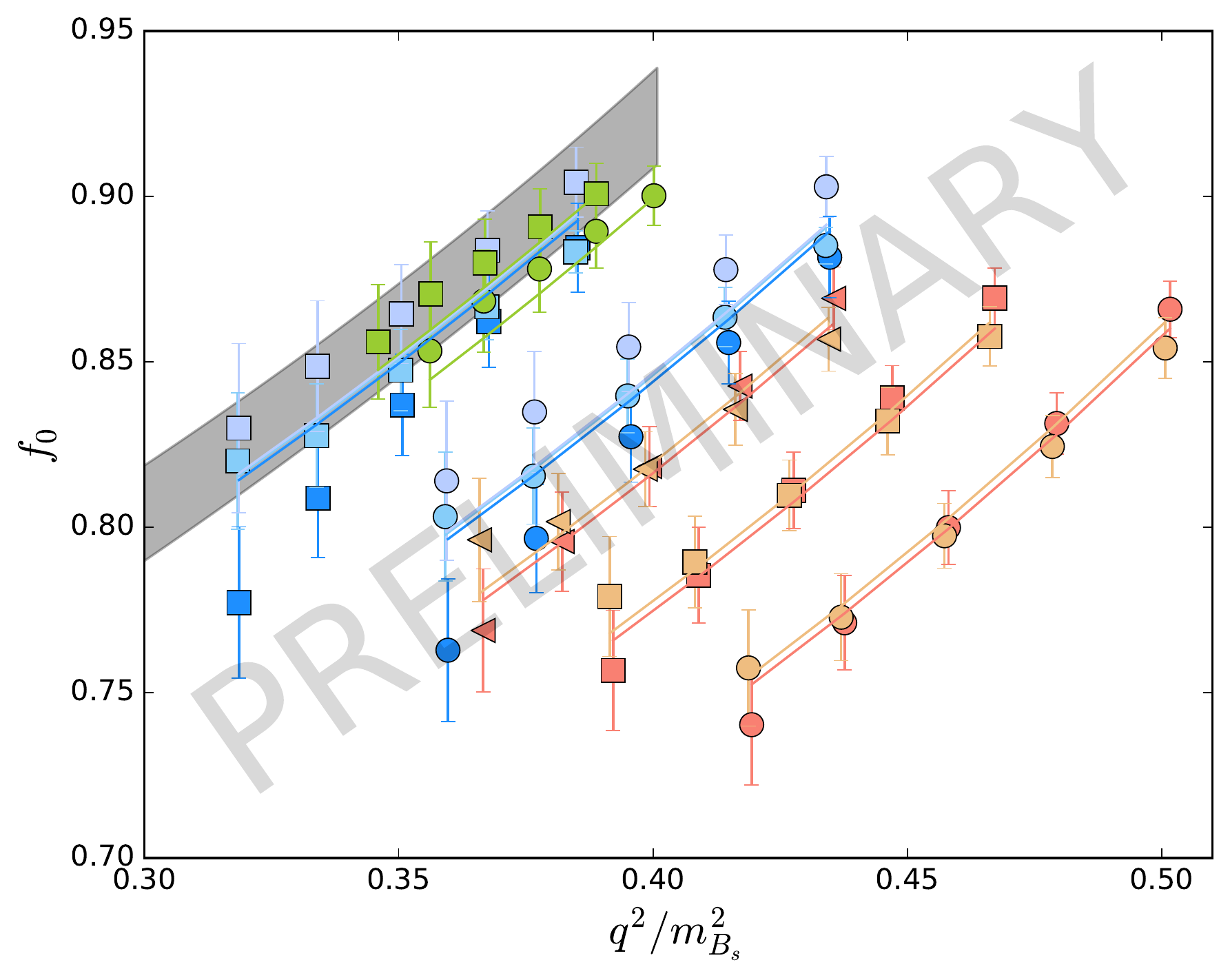}
    \caption{Semi-leptonic form factors $B_s \to D_s \ell\nu$ decays. The colored data points show our simulated data using charm-like masses to extra-/interpolate to the physical charm quark mass. Carrying out a global fit based on Eq.~(\ref{eq.BsDsfit}) we obtain the continuum limit (gray band) at physical quark masses.}
    \label{fig.BsDsdata}
\end{figure}

\begin{figure}[tb]
  \includegraphics[height=0.24\textheight]{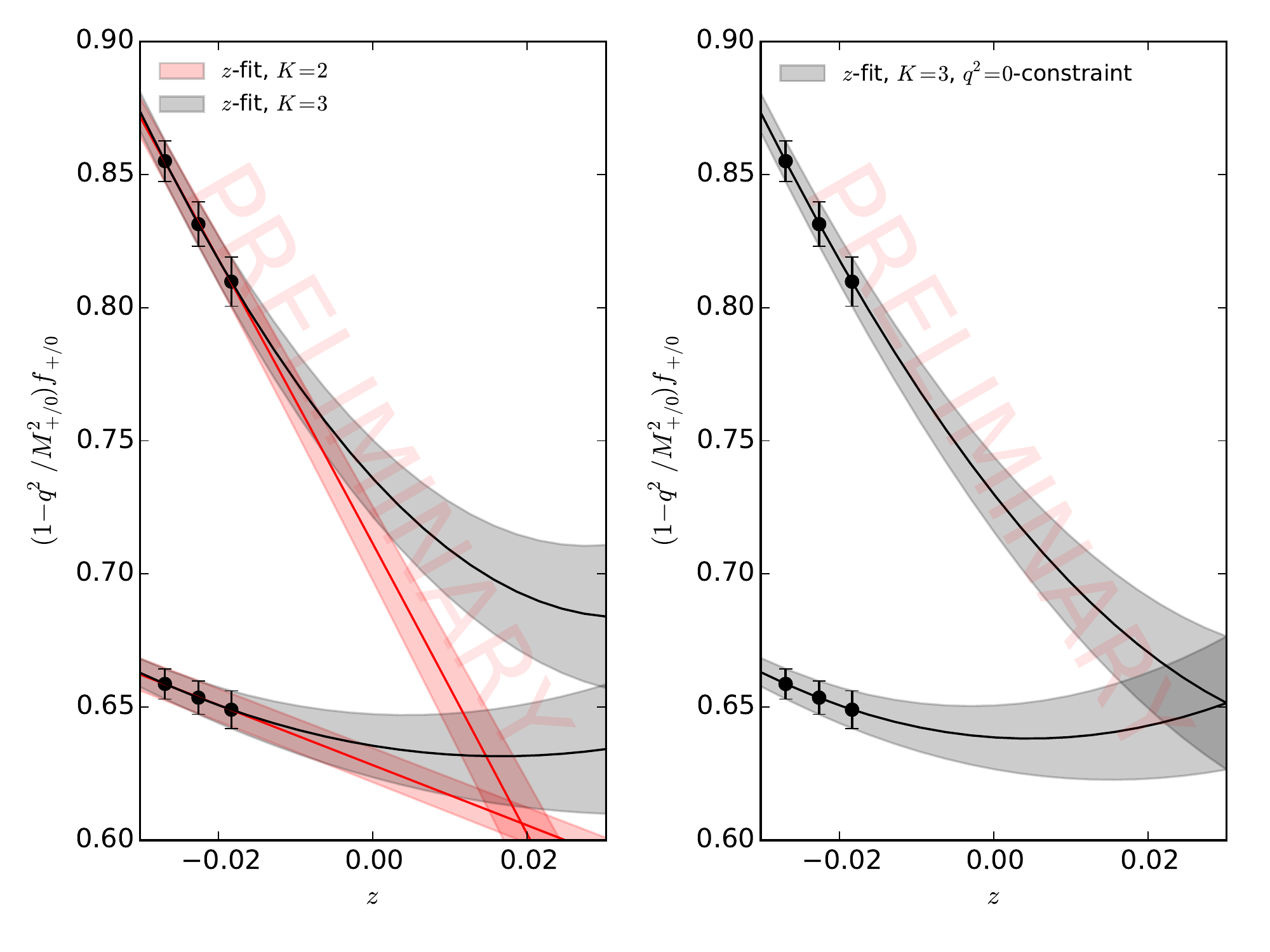}\hfill
  \includegraphics[height=0.24\textheight]{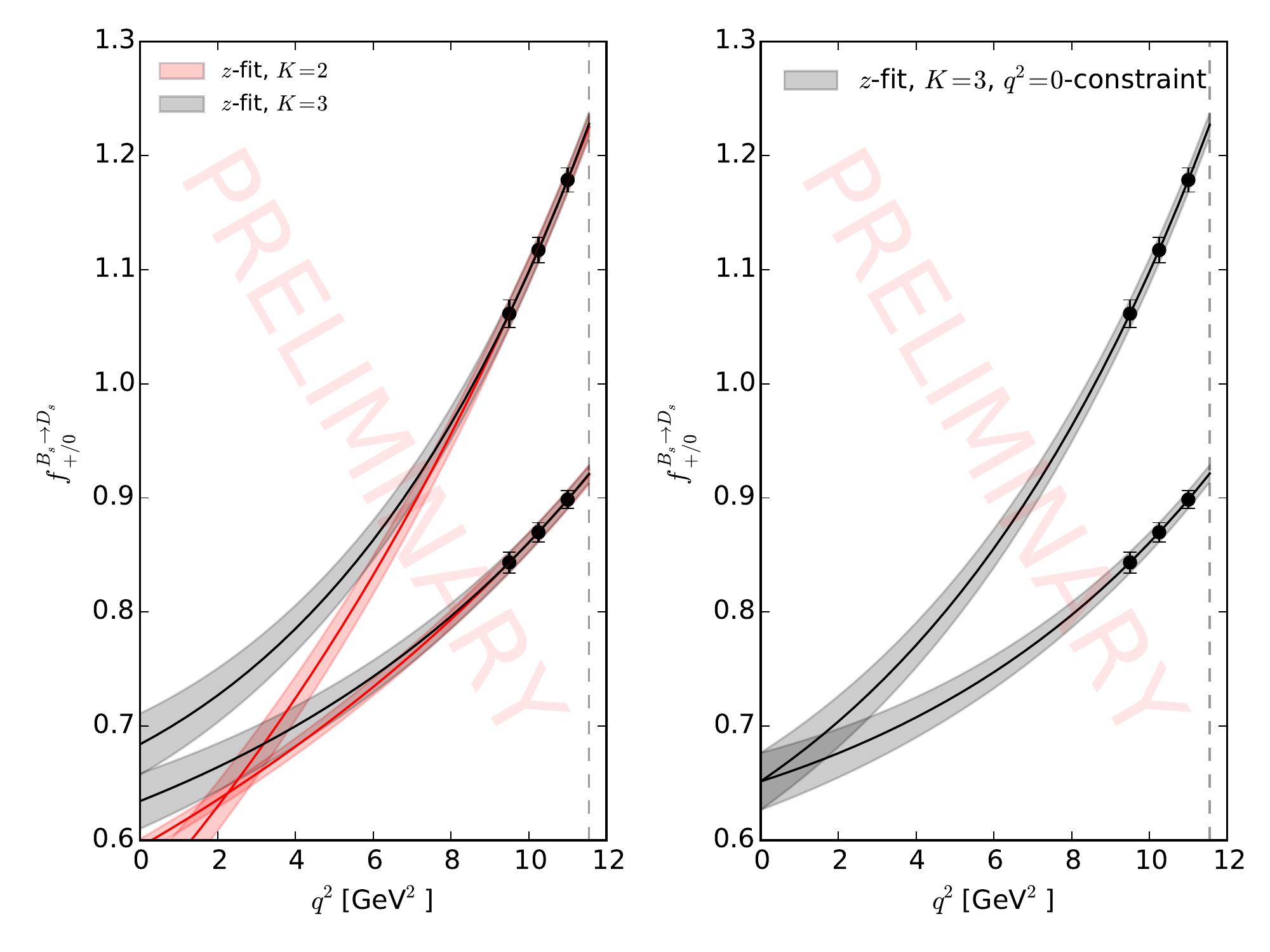}
  \caption{Kinematical extrapolation of the $B_s\to D_s\ell\nu$ form factors using the BCL $z$ expansion over the full $q^2$ range. On the left, the result are shown in units of the $z$ extrapolation, on the right in physical GeV\textsuperscript{2}.}
     \label{fig.BsDs_zfit}
\end{figure}

Our determination of $B_s\to D_s\ell\nu$ form factors follows steps analogous to those for $B_s\to K \ell \nu$ replacing the light daughter quark with a charm quark. We also choose to maintain the same parametrization in terms of $f_+$ and $f_0$.  However, the light quark mass only contributes to the sea-sector resulting in a mild (or flat) chiral extrapolation but, in addition, we need to perform an extra- or interpolation in the charm quark mass to obtain form factors for physical $D_s$ mesons \cite{Boyle:2017jwu}. This step is necessary because on the coarse ensembles we cannot directly simulate a physical charm quark mass with our choice of heavy domain wall action and on the medium and fine ensembles we choose to bracket the physical value of the charm quark mass. In Fig.~\ref{fig.BsDsdata} we show the form factors $f_+$  and $f_0$ obtained from our simulated charm quark masses i.e.~three charm-like masses are used to guide an extrapolation on the coarse ensembles (red symbols), while two charm-like masses enable an interpolation on the medium (blue symbols) and the fine (green symbols) ensembles. Again we obtain the result at physical quark masses performing a global fit, which for $B_s\to D_s\ell\nu$ is based on the ansatz
\begin{align}
  f(q^2,a, M_\pi, M_{D_s}) =    \frac{\alpha_0+\alpha_1 M_{D_s}+\alpha_2 a^2+\alpha_3 M_\pi^2}{1+\alpha_4 q^2/M_{B_s}^2},
  \label{eq.BsDsfit}
\end{align}  
to account for a dependence on the charm-quark mass, the lattice spacing, and the (sea) pion mass. The obtained continuum limit is shown by the gray error band.
As for $B_s\to K \ell \nu$, we are currently in the process of accounting for all systematic uncertainties and hence do not have a final and complete error budget, yet. Nevertheless we can proceed and perform a kinematical $z$ expansion using the BCL parametrization and show the current status using $K=2$ or 3 in Fig.~\ref{fig.BsDs_zfit}.

\section{Summary}
\label{Sec.summary} 
We reported updates on our calculation of semi-leptonic form factors for $B_s\to K \ell\nu$ and $B_s\to D_s\ell\nu$ decays. We are currently finalizing our error budgets. Our results will provide entirely independent determinations of the form factors so far also calculated by Atoui et al., HPQCD, and Fermilab/MILC \cite{Atoui:2013zza,Bouchard:2014ypa,Na:2015kha,Monahan:2017uby,Monahan:2018lzv,Bazavov:2019aom}. In addition our results will allow to extract R ratios for $B_s$ decays which may serve as proxy for corresponding $B$ decays and in addition allow for the determination of the ratio of CKM matrix elements $|V_{cb}/V_{ub}|$ in combination with an experimental measurement. Such an independent determination may also help to resolve present discrepancies betweeen inclusive and exclusive determinations.

\clearpage
{\small \paragraph{Acknowledgments} The authors thank our collaborators in the RBC and
UKQCD Collaborations for helpful discussions and suggestions.  Computations for
this work were performed on resources provided by the USQCD Collaboration,
funded by the Office of Science of the U.S.~Department of Energy, as well as on
computers at Columbia University and Brookhaven National Laboratory.
This work used the ARCHER UK National Supercomputing Service (\href{http://www.archer.ac.uk}{http://www.archer.ac.uk}).
Gauge field configurations on which our calculations are based were also generated
using the DiRAC Blue Gene Q system at the University of Edinburgh, part of the
DiRAC Facility; funded by BIS National E-infrastructure grant ST/K000411/1 and
STFC grants ST/H008845/1, ST/K005804/1 and ST/K005790/1.  This project has
received funding from the European Union's Horizon 2020 research and innovation
programme under the Marie Sk{\l}odowska-Curie grant agreement No 659322, the
European Research Council under the European Unions Seventh Framework Programme
(FP7/ 2007-2013) / ERC Grant agreement 279757, STFC grant ST/L000458/1 and ST/P000711/1, ST/P006760/1 through the DISCnet Centre for Doctoral Training. O.W.~acknowledges support by DOE grant DE-SC0010005. No new experimental data was generated for this research.}
%
%
{\small
\bibliography{../B_meson}
\bibliographystyle{apsrev4-1ow}
}



\end{document}